\documentclass[a4paper,amsmath,superscriptaddress,twocolumn,showkeys,showpacs,prb,aps, 10pt]{revtex4-1}
\usepackage{color}
\usepackage{graphicx}
\usepackage{tabularx}

\newcommand{\bk}{\mathbf{k}}
\newcommand{\bp}{\mathbf{p}}
\newcommand{\be}{\begin{equation}}
\newcommand{\ee}{\end{equation}}
\newcommand{\bea}{\begin{eqnarray}}
\newcommand{\eea}{\end{eqnarray}}
\newcommand{\ba}{\begin{eqnarray*}}
\newcommand{\ea}{\end{eqnarray*}}
\newcommand{\up}{\uparrow}
\newcommand{\down}{\downarrow}

\begin{document}

\title{Co adatoms on Cu surfaces: ballistic conductance and Kondo temperature}

\author{P. P. Baruselli}
\affiliation{SISSA, Via Bonomea 265, Trieste 34136, Italy}
\affiliation{CNR-IOM, Democritos Unit\'a di Trieste, Via Bonomea 265, Trieste 34136, Italy}
\affiliation{Institut f\"{u}r Theoretische Physik, Technische Universit\"{a}t Dresden, 01062 Dresden, Germany}

\author{R. Requist}
\affiliation{SISSA, Via Bonomea 265, Trieste 34136, Italy}

\author{A. Smogunov}
%\affiliation{CNR-IOM, Democritos Unit\'a di Trieste, Via Bonomea 265, Trieste 34136, Italy}
%\affiliation{ICTP, Strada Costiera 11, Trieste 34014, Italy}
%\affiliation{Voronezh State University, University Square 1, Voronezh 394006, Russia}
\affiliation{
{
Service de Physique de l'Etat Condens\'e (CNRS UMR 3680), DSM/IRAMIS/SPEC, CEA Saclay, 91191 Gif-sur-Yvette Cedex, France
}
}

\author{M. Fabrizio}
\affiliation{SISSA, Via Bonomea 265, Trieste 34136, Italy}
\affiliation{CNR-IOM, Democritos Unit\'a di Trieste, Via Bonomea 265, Trieste 34136, Italy}
%\affiliation{ICTP, Strada Costiera 11, Trieste 34014, Italy}

\author{E. Tosatti}
\affiliation{SISSA, Via Bonomea 265, Trieste 34136, Italy}
\affiliation{CNR-IOM, Democritos Unit\'a di Trieste, Via Bonomea 265, Trieste 34136, Italy}
\affiliation{ICTP, Strada Costiera 11, Trieste 34014, Italy}

\date{\today}

\begin{abstract}
The Kondo zero bias anomaly of Co adatoms probed by scanning tunneling microscopy is known to depend on the height of the tip above the surface, and this dependence is different on different low index Cu surfaces.  
% has different tip height dependence on different low index Cu surfaces.
% surfaces display different behavior depending on the Miller index of the surface.  
% As the tip approaches the adatom on the (100) surface, the Kondo temperature at first decreases and then increases.
On the (100) surface, the Kondo temperature first decreases then increases as the tip approaches the adatom, while on the (111) surface it is virtually unaffected.  % by the tip height.
% We study the Kondo effect of single Co adatoms on Cu(100) and Cu(111) surfaces, probed by an STM tip, using a DFT+NRG approach.  
% Density functional theory and numerical renormalization group (DFT+NRG) calculations correctly capture these trends.  
These trends are captured by combined density functional theory and numerical renormalization group (DFT+NRG) calculations.  The adatoms are found to be described by an $S=1$ Anderson model on both surfaces, and ab initio calculations help identify the symmetry of the active $d$ orbitals.  
% We show that the tip can affect the Kondo temperature on the Cu(100) surface, but not on the Cu(111) surface, in agreement with experimental results. 
We correctly reproduce the Fano lineshape of the zero bias anomaly for Co/Cu(100) in the tunneling regime but not in the contact regime, where it is probably dependent on the details of the tip and contact geometry. 
% We correctly reproduce the lineshape of the zero bias anomaly for Co/Cu(100) in the tunneling regime but not in the contact regime, where the geometrical details of the junction and nonequilibrium effects are likely to play a crucial role.  DFT+NRG also fails in describing 
% In agreement with prior results on other close-packed surfaces, t
The lineshape for Co/Cu(111) is presumably affected by the presence of surface states, which are not included in our method. 
% probably due to the presence of surface states, not taken into account by our method
% , and a mismatch in the symmetry of the % two 
% magnetic orbitals, different than the usually assumed $d_{z^2}$ type, necessitating a generalization of the Tersoff-Hamann approach.
% makes the standard Tersoff-Hamann approach fail. 
We also discuss the role of symmetry, which is preserved in our model scattering geometry  but most likely broken in experimental conditions.
%We believe that further work is needed to clarify these issues.

\end{abstract}

\maketitle

\section{Introduction}

Since the observation of zero bias anomalies (ZBA's) for Ce adatoms on silver \cite{schneider_ceag} and Co adatoms on gold\cite{madhavan_coau} by scanning tunneling microscopy (STM), the Kondo effect \cite{Kondo1964} of magnetic adatoms 
%on metallic surfaces 
has become a subject of extremely high interest.  STM measurements of the Kondo effect offer the possibility of achieving exquisite external control over a paradigmatic strongly correlated system.  % and a detailed understanding of electron transport in point contacts. 
% For more motivation can we give 1-2 reasons why it is of such high interest?
Several of the established Kondo systems have been reviewed in Ref.~\onlinecite{Ternes2009}.
%Ref.~\onlinecite{Ternes2009}.
%known: see Ref.~\onlinecite{Ternes2009} for a focused review.  
Despite the apparent simplicity of these systems, a full theoretical description is still lacking.  Indeed, great efforts using both ab initio and many-body approaches have been made, 
% cite these works here?
but several open issues still exist. 

The Kondo effect in magnetic adatoms has been successfully treated within  
an Anderson model approach \cite{anderson61} (mostly with a single impurity orbital of $d_{z^2}$ symmetry; only recently has the whole $d$ shell \cite{surer_cocuctqmc,jacob} 
% should we cite something of Jacob's work here? OK
been taken into account), but the role of the tip is still a subject of debate.  STM measurements have conventionally been performed in the tunneling regime where % the precise shape of 
the tip does not affect the results.
% is irrelevant.  
Recent works \cite{knorr2002, Neel2007, choi_cocu100, Vitali2008} have looked beyond tunneling measurements
% gone beyond this regime
to explore the contact regime,
% instead explored the contact regime, 
where the geometric details of the tip and its position above the adsorbate can affect the Kondo ZBA.  Ab initio calculations including tip-induced perturbations are needed to describe, for instance, the observed progression of the Kondo temperature as a function of tip height above a Co adatom on Cu(100) \cite{knorr2002,Neel2007,choi_cocu100}.
% an atomic scale 

% Recent works \cite{} have explored the contact regime as well; in some cases bonds are formed between the tip and adsorbate \cite{Venkataraman}.

STM conductance is usually calculated in the Tersoff-Hamann model \cite{tersoff_hamann,tersoff_hamann2}, in which the tip is described by a single $s$ orbital and current flows thanks to the coupling between the tip and nearby metallic states.  
% Generalizations of this model allow for conductance through $p$-like states \cite{nonsonoriuscitoatrovare}.  
%! introduce the notion of Fano lineshape earlier
However, the situation is more complicated in the presence of an adsorbate.  The interference between tunneling directly into the surface and tunneling via the adsorbate causes a ZBA in the STM conductance
\be
G(V)=G_{back} +\Delta G\frac{q^2+2qV/(k_BT_K)-1}{[q^2+1][(V/k_BT_K)^2+1]}
%G(V)=G_{back} +\Delta G\frac{[q+V/(k_BT_K)]^2}{[q^2+1][(V/k_BT_K)^2+1]}
\label{eq:zba}
\ee
with a characteristic Fano lineshape \cite{fano}, the fingerprint of the Kondo effect;
here we introduce a $1/(q^2+1)$ factor in such a way that  $\Delta G/G_{back}$
represents the signal to background ratio, which is experimentally found to be on the order of 10-30$\%$.
The parameter $q$ describes the shape of the ZBA ($q=0$ corresponds to a minimum, $q=\pm\infty$ to a maximum, while intermediate values give rise to asymmetric lineshapes), while $T_K$, the Kondo temperature, is proportional to its width ($k_B$ is the Boltzmann constant).

%Electrical conductance through magnetic adatoms can be calculated with the Landauer-B\"uttiker formalism \cite{landauer}.
% If a magnetic adatom measured by STM is treated as a nanocontact, the 
% ballistic conductance
The full STM-adatom geometry can alternatively be viewed as a nanocontact and the electrical conductance calculated within the Landauer-B\"uttiker formalism \cite{landauer}.  % However, density functional theory (DFT) calculations in the standard semi-local approximations do not give the correct ballistic scattering matrix because they do not account for the many-body correlations responsible for Kondo ZBA's. 
But when Kondo correlations are present, the ballistic scattering matrix cannot be obtained from density functional theory (DFT) in the standard semi-local approximations because the latter do not include the many-body correlations responsible for ZBA's.  
% for exchange-correlation potential 
DFT calculations are nevertheless indispensable in singling out the relevant adsorption geometries and electronic degrees of freedom.  % support specific impurity spin models \cite{} and 

% However, density functional theory (DFT) calculations in the standard semi-local approximations 
% for exchange-correlation potential 
% do not give the correct ballistic scattering matrix because they do not account for the many-body Kondo correlations responsible for the observed ZBA's.
% To achieve quantitative ab initio modeling of Kondo ZBA's, recent works \cite{lucignano,baruselli,requist} have developed a scheme marrying DFT and NRG via an intermediate Anderson impurity model.  

To model Kondo ZBA's from first principles, recent works \cite{lucignano,requist_no_kondo,prb_nanotubi} have developed a scheme to quantitatively join DFT and many-body calculations via an intermediate Anderson impurity model (AIM).  The model parameters are determined by matching the mean-field scattering matrix of the AIM to the ballistic scattering matrix of a spin-polarized DFT calculation.  Observables are then obtained by solving the AIM with the many-body numerical renormalization group (NRG) method.
% The mean-field scattering matrix of the AIM is matched to the scattering matrix of a spin-polarized DFT calculation to determine the model parameters.  Once the model parameters are fixed, relevant physical observables are calculated with the many-body numerical renormalization group (NRG) method.  
% This DFT+NRG scheme predicted a Kondo ZBA for the nitric oxide molecule adsorbed on the Au(111) surface which was verified in experiment \cite{requist_no_kondo}, and similar predictions have been made for Co impurities and vacancies in carbon nanotubes \cite{Baruselli_physE,baruselli_prl2012,prb_nanotubi}.
This DFT+NRG scheme has successfully predicted the Kondo ZBA of nitric oxide adsorbed on the Au(111) surface, albeit underestimating the experimental Kondo temperature \cite{requist_no_kondo}.
%This DFT+NRG scheme has been experimentally validated in an STM nanocontact geometry where it was successful in predicting a Kondo ZBA for the nitric oxide molecule adsorbed on the Au(111) surface, albeit underestimating the experimental Kondo temperature \cite{requist_no_kondo}, 
Predictions have also been made for Co impurities and vacancies in carbon nanotubes \cite{Baruselli_physE,baruselli_prl2012,prb_nanotubi}.  Different approaches that incorporate many-body correlations into first principles calculations through dynamical mean-field theory have also been proposed \cite{thygesen, jacob}.
% Another approach is to incorporate many-body correlations into first principles calculations through dynamical mean-field theory \cite{thygesen, jacob}.

Two main approaches have been adopted to calculate the Fano parameter $q$ in Eq.~(\ref{eq:zba}). 

The first is the ``two-path model'' \cite{schiller,castroneto_cocu111, madhavan_coau_prb, knorr2002, plihal_gadzuk,morr_2010}, in which the tip is coupled to the adatom $d_{z^2}$ orbital via hopping $t_{\bp d}$ and to the surface via $t_{\bp\bk}$, where $\bk$ denotes conduction states of the surface  and $\bp$ the states of the tip, giving the expression
\be\label{q_t1t2}
q=\frac{t_{\bp d}+\sum_\bk t_{\bp\bk} V_{\bk d}\Re G_\bk}{\sum_\bk t_{\bp\bk}V_{\bk d} \Im G_\bk}
\rightarrow \frac{t_2+t_1 V_d \Re G(0)}{t_1 V_d \Im G(0)},
\ee
where $V_{\bk d}$ are matrix elements between the $d$ orbital and the surface and $G_\bk$ is the Green's function of the clean surface.  By considering $t_{\bp\bk}$, $t_{\bp d}$ and $V_{\bk d}$ to be energy and momentum-independent, and calling them respectively $t_1$, $t_2$ and $V_d$, the right-hand side of Eq.~(\ref{q_t1t2}) is obtained, where only the surface Green's function $G(0)$ at the Fermi energy appears.

The second approach consists in neglecting the coupling of the tip to $d$ states, i.e.~$t_2 \equiv t_{\bp d}=0$,\cite{merino_gunnarsson,ujsaghy} so that just the density of states of the metal is probed by STM via $t_1 \equiv t_{\bp\bk}$, assumed for simplicity to be momentum-independent; this leads to the Fano parameter
\be
q=\frac{\sum_\bk t_{\bp\bk} V_{\bk d}\Re G_\bk}{\sum_\bk t_{\bp\bk}V_{\bk d} \Im G_\bk}
\rightarrow \frac{\Re G(0)}{\Im G(0)},
\ee
which is zero for a particle-hole symmetric band 
%$\Re G(0)=0$ 
if the the momentum dependence of the matrix elements can be neglected. 
This second approach is justified for Co/Au by the fact that the experimentally observed Fano resonance is independent of tip height.\cite{madhavan_coau_prb}  
% If $t_2$ was not negligible, $q$ would have non-negligible $z$ dependence because $t_{1}$ and $t_{2}$ are generally position dependent parameters $t_{1}(R,z)$ and $t_{2}(R,z)$ with different behavior as a function of the tip position $(R,z)$ ($z$ being the height above the adatom and $R$ the lateral displacement).  
If $t_2$ were not negligible, $q$ would be expected to have non-negligible $z$ dependence because $t_{1}$ and $t_{2}$ generally have different dependence %depend 
on the tip position $\mathbf{R}=(R,\varphi,z)$ ($z$ being the height above the adatom and $R$ the lateral displacement in the angular direction $\varphi$).  The $R$ dependence of $q$ has been observed, % add citation
but it is mainly a consequence of probing variations of the surface Green's function at different positions.

In this paper, we use our DFT+NRG scheme to study how the Kondo temperature and Fano line shape are affected by the location of the tip in two experimentally well-characterized cases: single Co impurities on Cu(100) and Cu(111) surfaces \cite{knorr2002, Neel2007, choi_cocu100, Vitali2008}.    
Experimentally, it is found that on the Cu(100) surface Co adatoms show a ZBA with $T_K=88$K and $q=1.13$; upon moving the tip laterally $q$ decreases down to 0.6 \cite{knorr2002}. 
When the tip approaches the Co adatom, $T_K$ increases to 700K  and $q$ to $\sim 70$ in one experiment \cite{choi_cocu100} and $T_K$=150K and $q\sim2$ in another \cite{Neel2007}. 
On the Cu(111) surface, $T_K=54$K and $q=0.18$; upon moving the tip laterally, $q$ decreases to $\sim 0$;\cite{knorr2002} when the tip approaches the adatom, both $T_K$ and $q$ are unaffected \cite{Vitali2008}.

Our main results can be summarized as follows.  Spin polarized DFT calculations show that on both surfaces the spin state of Co is $S=1$, 
% to a good approximation
each of two magnetic $d$ orbitals contributing approximately one Bohr magneton.  On the (100) surface these orbitals are inequivalent; 
%   --- $d_{z^2}$ and $d_{x^2-y^2}$ --- 
% when performing NRG calculations, 
the $d_{z^2}$ orbital is found to have a much higher Kondo temperature than the $d_{x^2-y^2}$ orbital.  
The effect of the tip is to increase the hybridization of orbital $d_{z^2}$ by pushing the adatom into the surface, as well as itself providing another source of hybridization.
% When an STM tip approaches the adatom, it acts as an additional source of hybridization, thus effectively increasing the Kondo temperature.
On the (111) surface, the two magnetic orbitals are degenerate, and their hybridization does not increase as the STM tip gets closer, due to symmetry and structural reasons.  Consequently, the Kondo temperature does not vary appreciably all the way from the tunneling regime to the contact regime.
On both surfaces a precise determination of the Kondo temperature, which depends exponentially on the parameters of the AIM, is beyond the capabilities of our method.  Both the Kondo temperature and the Fano parameter $q$ are affected by numerous fine details, comprising the electronic structure of the surface and adatom, the details of the tip--adatom--surface nanocontact, in particular, how strongly symmetry is broken by the tip, and possibly non-equilibrium and multi-orbital (beyond 2) effects.  
% The latter are a potential subject for future research.

The paper is structured as follows. 
In Section \ref{sec_ourapproach} we describe our method; 
in Sections \ref{sec_cocu100} and \ref{sec_cocu111} we present our results for Co/Cu(100) and Co/Cu(111), respectively; and in Section \ref{sec_conclusions} we discuss the conclusions of our work.

\section{Overview of the method}\label{sec_ourapproach}
We have employed, with a few simplifications, the method presented in Refs.~\onlinecite{lucignano, Baruselli_physE, baruselli_prl2012, prb_nanotubi,requist_no_kondo}, to which we refer for further details. 

\begin{figure}[htb]
\centering
 \includegraphics[scale=0.28]{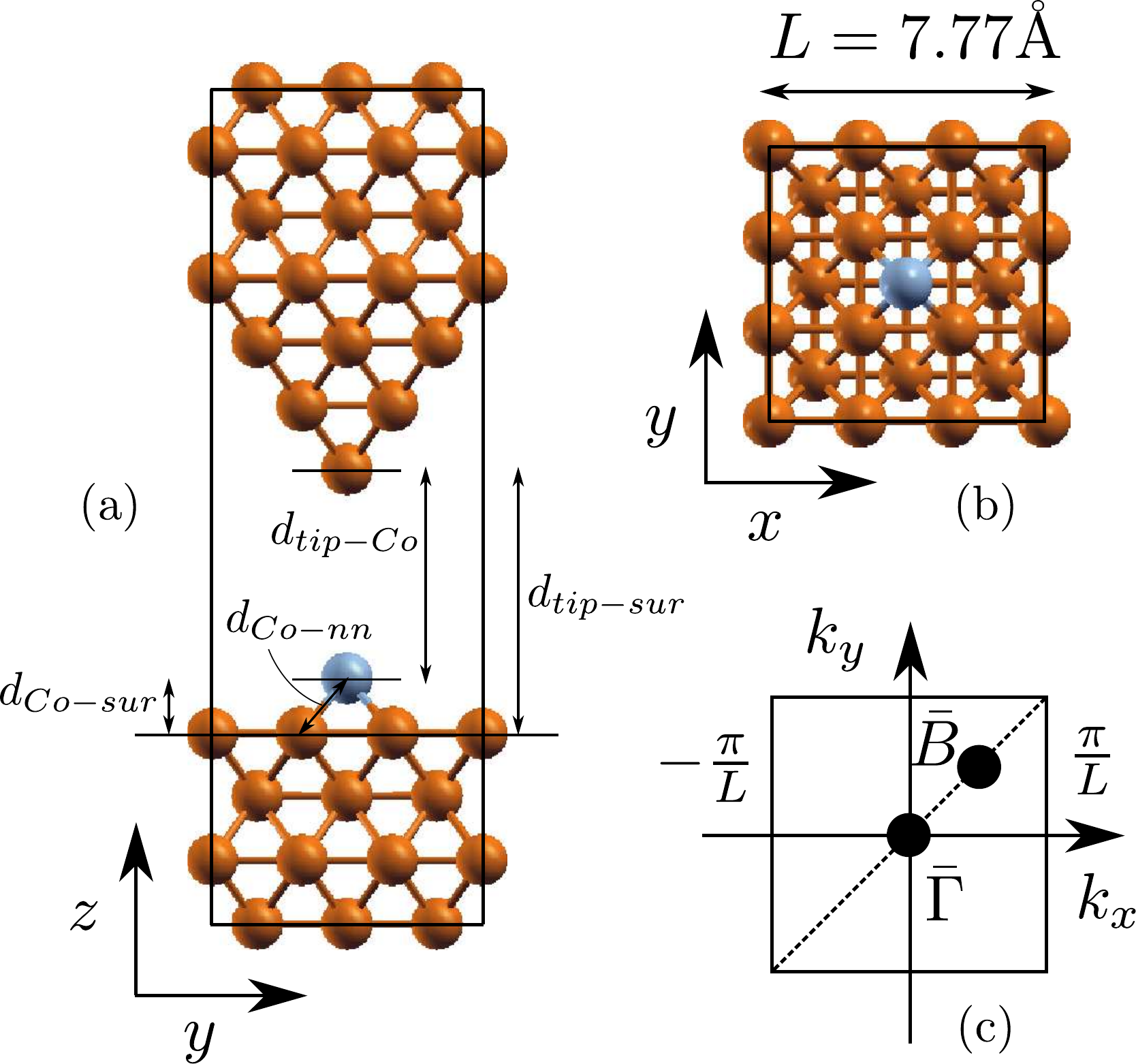}% \includegraphics[width=\textwidth]{../imm/cocu100cellb.png}
\caption{Unit cell of the scattering region for Co/Cu(100) for $d_{tip-sur}=7.60\text{\AA}$ (a) in the $yz$ plane and (b) in the $xy$ plane. (c) Brillouin zone for the supercell in the $xy$ plane. % associated with the supercell.
}\label{cocu100cell}
\end{figure}

First we perform a self-consistent, fully relaxed calculation of the electronic properties of the scattering region (as shown in Figs.~\ref{cocu100cell} and \ref{cocu111cell}) by density functional theory. This is constituted by a $3\times 3$ Cu supercell in the $xy$ plane with a Co coverage of 1/9; in the $z$ direction we use 8 Cu layers plus a ``pyramid'' of 5 Cu atoms to simulate the STM tip for the (100) surface; for the (111) surface, we use 7 layers and a 4-atom pyramid. At this coverage the interaction between periodic replicas of the adatom is small.
The calculations are carried out with the standard plane-wave package QUANTUM ESPRESSO \cite{QE-2009} using the generalized gradient approximation (GGA) to the exchange-correlation functional in the parametrization of Perdew, Burke and Ernzerhof ~\cite{pbe}. The plane wave cut-offs are 30 Ry and 300 Ry for the wave functions and charge density, respectively. Integration over the two-dimensional Brillouin zone is accomplished using a $6\times 6$ grid of $\bk$-points and a smearing parameter of $10$ mRy.

After obtaining the self-consistent electronic structure, the conductance in the $z$ direction is calculated 
%, perpendicular to the surface, 
using the PWCOND routine.\cite{2004Smogunov-PRB}
% , which is part of the Quantum-ESPRESSO package. 
Scattering eigenchannels and eigenvalues depend on the transverse momentum $\bk_{xy}$; hence, we restrict our conductance calculations to the single most representative
% a single as representative as possible 
$\bk_{xy}$ points: $\bar{B}=\frac{\pi}{L}(\frac{1}{2},\frac{1}{2})$ for the (100) surface, and $\bar{K}=\frac{\pi}{L}(\frac{2}{3},0)$ for the (111) surface. 
This procedure introduces small systematic errors in the estimation of parameters but has the advantage of keeping the computational effort low.  We verified that the error with respect to a more accurate calculation with $5\times 5$ $\bk_{xy}$ points is less than a few percent.
In the above expressions, $L=$7.77\AA~is the length of the supercell in the $x$ and $y$ directions, set to three times the equilibrium nearest-neighbor distance for bulk Cu for our pseudopotential, 2.59\AA, slightly larger than the experimental value 2.56\AA.

With the knowledge of the scattering eigenvalues $t_n$, it is possible to compute the 
% () 
energy-dependent Fano factor\cite{fano_factor}:
\be
F=\frac{\sum_n t_n(1-t_n)}{\sum_n t_n}
\ee
which is experimentally accessible through noise measurements.\cite{giant_fano_factor,shot_noise_fano} 

In the final step, an AIM is built in such a way that it reproduces the DFT scattering phase shifts as closely as possible when solved in the Hartree-Fock (HF) approximation.  
The Kondo temperature can be estimated after the AIM is solved by numerical renormalization group (NRG) \cite{wilson, bulla08}. 
In contrast to Refs.~\onlinecite{lucignano, prb_nanotubi} where a phase-shift analysis was performed to determine the Fano parameter $q$, here, due to the additional complication of the dependence on transverse momentum and lack of even/odd symmetry along the $z$ direction, the lineshape is inferred %guessed 
directly at the DFT level by looking at the energy-dependent transmission eigenvalues,
and fitting them with a Fano lineshape.

This procedure is repeated for a series of tip--surface distances to show how the Kondo temperature and lineshape vary in going from the tunneling to contact regime.

\section{Co/Cu(100)}\label{sec_cocu100}
In this section we present results for the Co/Cu(100) system, which is found experimentally to have $T_K=88$ K and $q=1.13$ in the tunneling regime\cite{knorr2002}. These values are modified in the contact regime, where $T_K$ grows up to 700 K and $q$ up to $\sim$70, \cite{choi_cocu100}.  Ref.~\onlinecite{Neel2007} found that $q$ does not go beyond 2 -- a discrepancy which is probably due to the %slightly 
different nanocontact geometries in different experimental conditions.

\subsection{DFT results}
We find that the Co adatom adsorbs in the hollow position, with 4 nearest-neighbor Cu atoms. In this configuration the symmetry group is $C_{4v}$, and Co $3d$ orbitals are split into 3 singlets ($d_{z^2}$ with symmetry A1, $d_{x^2-y^2}$ with symmetry B1, $d_{xy}$ with symmetry B2) and a doublet ($d_{xz}$ and $d_{yz}$ with symmetry E); the $4s$ orbital has A1 symmetry.  In our scattering geometry the tip is built so as not to lower the $C_{4v}$ symmetry of the adatom plus surface system.  We discuss possible consequences of this approximation later, since symmetry is not preserved in real experiments. 

GGA calculations, both with and without the tip, show the presence of two magnetic orbitals, $d_{z^2}$ and $d_{x^2-y^2}$, in agreement with Ref.~\onlinecite{huang_carter_2008}, while the $4s$ orbital is highly hybridized, almost spin-unpolarized and singly occupied; the electronic configuration is thus $3d^84s^1$, and the magnetic moment is close to $2\mu_B$, in agreement with, for example, Ref. \onlinecite{polok_cocu}.  Table~\ref{tab_cocu100} 
%! Let's remove some columns - maybe the epsilons, U12 and J
reports some structural and electronic data for this system at different tip--surface separations.  Structural data are in good agreement with Ref.~\onlinecite{huang_cocu100}.  
%!
Table~\ref{tab_cocu100} covers the approximately-known range of experimental tip--surface separations.  The tip--surface separation could not be further reduced because already at the smallest value reported in Table~\ref{tab_cocu100}, $d_{tip-sur}=4.12 \text{\AA}$, $\epsilon_{z^2}$ and $U_{z^2}$ are affected by large errors.
% It was not possible to reduce the tip--surface separation beyond the shortest value reported in Table~\ref{tab_cocu100}, $d_{tip-sur}=4.12 \text{\AA}$, $\epsilon_{z^2}$ and $U_{z^2}$ are affected by a large error because 
The source of error is the breakdown of our procedure for estimating parameters when the occupation of the $d_{z^2}$ orbital grows significantly greater than 1 
% orbital starts to become occupied significantly more than 1, 
and part of the polarization is transferred to the $d_{xz}$ and $d_{yz}$ orbitals. 
This is expected since with increased coordination Co approaches the configuration it has as a bulk impurity, where the polarization is shared by all $d$ orbitals\cite{koritar_cocu111}.  

\begin{figure}[ptb]
 \includegraphics[width=0.48\textwidth]{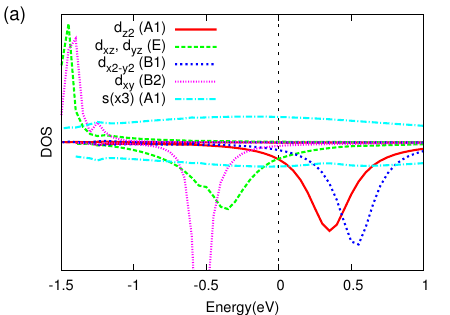}
 \includegraphics[width=0.48\textwidth]{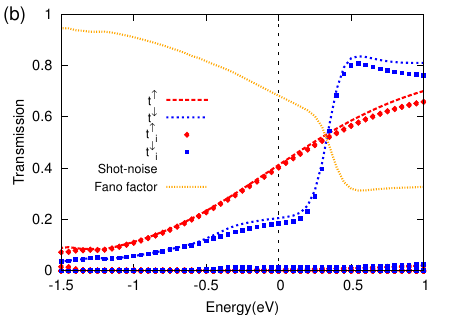}
\caption{Co/Cu(100): (a) Spin-polarized density of states { at Co $3d$ and $4s$ atomic orbitals constructed from scattering states} at $\bk_{x,y}=\bar{B}= \frac{\pi}{L}(\frac{1}{2},\frac{1}{2})$ %for $d_{tip-sur}=5.58\text{\AA}$ 
(positive DOS=spin up, negative DOS=spin down); 
(b) { $t^{\uparrow}$ and $t^{\downarrow}$ components of the DFT transmission, 
transmission eigenvalues $t^{\uparrow}_i$ and $t^{\downarrow}_i$} ($\sum_i t^{\uparrow}_i=t^\uparrow$,  
$\sum_i t^{\downarrow}_i=t^\downarrow$) and the shot noise Fano factor.  
Energies are with respect to the Fermi energy.}\label{cocu100dos4}
\end{figure}

Figure~\ref{cocu100dos4} shows representative density of states and transmission eigenvalues for $d_{tip-sur}=5.58\text{\AA}$.  The interference between the $s$ and $d_{z^2}$ orbitals in the down-spin A1 channel gives a Fano lineshape with $q\sim1$ centered at an energy around $\epsilon_{d_{z^2}}^\downarrow \sim 0.5$ eV; all other channels give a negligible contribution to the total conductance.

\begin{table*}
\centering
\begin{ruledtabular}
\begin{tabular}{ccccccccccccccccccc}
$d_{tip-sur}$ & $d_{tip-Co}$ & $d_{Co-sur}$ & $d_{Co-nn}$ & $m$ & $\epsilon_{z^2}$ & $U_{z^2}$ & $\Gamma_{z^2}$ & $\epsilon_{x^2-y^2}$ & $U_{x^2-y^2}$ & $\Gamma_{x^2-y^2}$ & $U_{12}$ & $J$ & $T_{K,z^2}$ & $T_{K,x^2-y^2}$ & $q_{z^2}$ & $g$ & $F$ \\\colrule
%6.00	pp
7.60 	&6.01 	&1.59 	&2.43 &	2.08	&
-4.83&3.06&0.183&-4.89&3.24&0.147&1.56&1.28&340&50&1.20&0.01&0.99\\	
%4.00	
5.58 &	3.91 	&1.67 &	2.47 &	2.10	&
-4.67&2.98&0.165&-4.80&3.26&0.130	&1.50&1.29&290&10&1.19&0.31&0.68&\\
%3.60	
5.17 &	3.41 &	1.76 &	2.51 &	2.10&	
-4.54&2.91&0.170&-4.67&3.23&0.116&1.44&1.30&410&3&0.76&0.61&0.53\\
%3.15	
4.73 &	2.84 &	1.89 &	2.56 &	2.08&	
-4.47&2.86&0.195&-4.60&3.23&0.101&1.40&1.32&600&0.5&0.09&1.06&0.42\\
%2.90	
4.48 	&2.61 	&1.87 &2.54 &	2.05&
-4.58&2.88&0.211&-4.65&3.22&0.104&1.40&1.32&1000&1&0.03&1.27&0.38\\
%2.52	
4.12 	&2.42 &	1.70 &	2.48 &	1.96&	
-5.47$^*$ &3.36$^*$ &0.232&-5.23&3.22&0.122&1.63&1.33&1100&25&0.01&1.36&0.34
\end{tabular}
\end{ruledtabular}
\caption{Parameters for Co/Cu(100) at different tip--surface separations $d_{tip-sur}$:
the distance $d_{tip-Co}$ between the tip and the Co adatom, 
$d_{Co-sur}$ between the Co adatom and the surface, 
$d_{Co-nn}$ between the Co adatom and its 4 nearest neighbors, 
the total magnetization $m$ of the unit cell in $\mu_B$,
the on-site energies $\epsilon_{z^2}$ and $\epsilon_{x^2-y^2}$, 
the Hubbard repulsions $U_{z^2}$ and $U_{x^2-y^2}$, 
the hybridizations $\Gamma_{z^2}$ and $\Gamma_{x^2-y^2}$, 
the inter-orbital Hubbard repulsion $U_{12}$, 
the Hund exchange $J$, 
the calculated Kondo temperatures $T_{K,z^2}$ and $T_{K,x^2-y^2}$ in $K$,
% from the NRG solution of Eq.~\ref{h_ad_sur100},
the Fano parameter $q_{z^2}$,
% ($q_{x^2-y^2}$, not shown, is always $\gg 1$), 
the DFT conductance $g=G/G_0=g^\uparrow+g^\downarrow$ and Fano factor $F$ at the Fermi energy; distances are in \mbox{\AA} and energies in eV.  
$^*$Here, $\epsilon_{z^2}$ and $U_{z^2}$ have large errors (see text).
}\label{tab_cocu100}
\end{table*}


\subsection{Anderson model}\label{sec_aim_cocu100}
In building an effective AIM, only the $d_{z^2}$ and $d_{x^2-y^2}$ magnetic orbitals are retained, each coupled to a linear combination of conduction states with the same symmetry, A1 and B1, respectively. Within this approximation, the adatom--surface Hamiltonian is
\bea\label{h_ad_sur100}
H_{ad-sur}&=&\sum_{\substack{\bk\sigma\\i=A1,B1}}\epsilon_{\bk i}c^\dagger_{\bk i\sigma}c_{\bk i\sigma}+\nonumber\\ &&+\sum_{\substack{\bk\sigma\\i=A1,B1}} V_{\bk i} (c^\dagger_{\bk i\sigma} d_{i \sigma}+d_{i\sigma}^\dagger c_{\bk i\sigma} )+\nonumber\\
&&+\sum_{i=A1,B1}(\epsilon_i n_i +U_i n_i^\uparrow n_i^\downarrow)+ \nonumber\\
&&+ U_{12} n_{z^2}n_{x^2-y^2}-J\boldsymbol{S}_{z^2}\cdot \boldsymbol{S}_{x^2-y^2},
\eea
where we have introduced the single-particle energies $\epsilon_{\bk i}$ and fermionic operators $c_{\bk i\sigma}$ and $c_{\bk i\sigma}^{\dag}$ associated with conduction states with momentum $\bk$, symmetry $i$, and spin $\sigma$, the on-site energies $\epsilon_i$ and fermionic operators $d_{i\sigma}$ and $d_{i\sigma}^{\dag}$ associated with impurity orbitals ($d_{A1}\equiv d_{z^2}$, $d_{B1}\equiv d_{x^2-y^2}$), the hopping elements $V_{\bk i}$ between a conduction state 
%with momentum $\bk$ and symmetry $i$ 
and $d$ orbital, the on-site Hubbard repulsion $U_i$ for $d_{z^2}$ and $d_{x^2-y^2}$ orbitals, the inter-orbital Hubbard repulsion $U_{12}$ between $d_{z^2}$ and $d_{x^2-y^2}$, and the Hund exchange $J>0$ between $d_{z^2}$ and $d_{x^2-y^2}$. We denote $n_i^\sigma=c^\dagger_{i\sigma} c_{i\sigma}$, $n_i=\sum_\sigma n_i^\sigma$, $\boldsymbol{S}_{i}= \frac{1}{2}\sum_{\mu\nu}d^\dagger_{i\mu}\boldsymbol{\sigma}_{\mu\nu} d_{i\nu}$, where $\boldsymbol{\sigma}$ is the vector of Pauli matrices $\boldsymbol{\sigma}=(\sigma_x, \sigma_y, \sigma_z)$.

A generalized Tersoff-Hamann 
model is used for the interaction of the tip with the surface and the $d$ orbitals on the adatom:
\bea\label{h_tip_100}
H_{tip}&=&\sum_{\substack{\bp\sigma\\i=A1,B1}} \epsilon_{\bp i} c^\dagger_{\bp i\sigma} c_{\bp i\sigma} +\nonumber\\ &&+\sum_{\substack{\bp\sigma\\i=A1,B1}} t_{2i} (c^\dagger_{\bp i\sigma} d_{i\sigma}+d^\dagger_{i\sigma} c_{\bp i\sigma})+\nonumber\\
&&+\sum_{\substack{\bp\bk\sigma\\i=A1,B1}} t_{1i} (c^\dagger_{\bp i\sigma} c_{\bk i\sigma} + c^\dagger_{\bk i\sigma} c_{\bp i\sigma}), 
\eea
where $\epsilon_{\bp i}$ and $c_{\bp i\sigma}$ denote the single-particle energies and destruction operators associated with conduction states of the tip with momentum $\bp$, symmetry $i$, and spin $\sigma$. 
%Parameters $t_{1A1}$ and $t_{1B1}$ describe a tip--surface hopping, supposed to be momentum-independent, respectively for A1 and B1 symmetry; parameters $t_{2A1}$ and $t_{2B1}$ describe tunneling from the tip into the $d$ states. 
The usual approach only takes into account the $A1$ symmetry channel, i.e.~the tip--surface hopping $t_{1A1}$ and $t_{2A1}$, which is motivated by the fact the apex atom of the tip (Cu in this case) has a single $s$ orbital at the Fermi energy. In contrast, we allow for conductance through channels with different symmetry. 
Since our geometry preserves the symmetry of the Co $d$ states, the total conductance $g_{tot}\equiv G_{tot}/G_0$, where $G_0=e^2/h$ is the quantum of conductance, is the sum of the conductance from A1 and B1 channels (we ignore all other symmetry channels as they do not contribute to the ZBA):
\be\label{gtot100}
g_{tot}=\sum_{i=A1,A2,B1,B2,E}g_i\simeq g_{A1}+g_{B1}.
\ee
We are aware that symmetry is not preserved in real experiments, because the tip cannot be expected to have the ideal pyramid-like shape we have assumed; however, it is a good starting point for studying the problem.  Should symmetry be broken, channels with different symmetries would start interfering, leading to a modified lineshape. However, we believe the disagreement we find with the experimentally determined $q$ is only be partially due to this approximation: since a single Kondo temperature is relevant (see next subsection), the main effect of interference would be to slightly modify the hopping parameters from the tip to the surface and the adsorbate, only weakly affecting our estimate of $q$.

We now introduce the hybridization functions $\Gamma_{i}^s(\epsilon)$ due to the coupling with the surface,  $\Gamma_{i}^t(\epsilon)$ due to the coupling with the tip, and the total $\Gamma_{i}(\epsilon)$:
\bea
\Gamma_{i}^s(\epsilon)&=&\pi\sum_\bk \delta (\epsilon-\epsilon_{\bk i}) V_{\bk i}^2\rightarrow\pi\rho_{si} V_{i}^2,\\
\Gamma_{i}^t(\epsilon)&=&\pi\sum_\bk \delta (\epsilon-\epsilon_{\bk i}) t_{\bk i}^2\rightarrow\pi\rho_{ti} t_{2i}^2,\\
\Gamma_{i}(\epsilon)&=&\Gamma_{i}^s(\epsilon) + \Gamma_{i}^t(\epsilon), \hspace{10pt} i =d_{z^2}, d_{x^2-y^2},
\eea
where the expressions $\pi\rho_s V_{i}^2$ and $\pi\rho_t t_{2i}^2$ are valid as long as we assume energy independent quantities ($\rho_{si}$ and $\rho_{ti}$ denote the density of states of symmetry $i$ at the Fermi energy for the surface and the tip, respectively). Our method only allows us to infer the total $\Gamma_i(\epsilon)$, which we assume to be energy-independent, and simply call $\Gamma_i$. Also, we assume that the density of states of conduction electrons is flat, and extends from $-D_{i}$ to $D_i$:
\be
\sum_\bk \delta(\epsilon-\epsilon_\bk)=\frac{\theta(D_i-|\epsilon|)}{2D_i}, \hspace{10pt} i=A1,B1 {,}
\ee
and we take $D_{A1}=D_{B1}=7$eV as estimated from the density of states at the Fermi energy of $s$ electrons for Cu atoms on the clean surface.

In principle, the $4s$ orbital of the Co atom should also be taken into account:
\bea\label{h_s}
%H=t_{d} \sum_p  (c^\dagger_{p} d_{z^2} +d^\dagger_{z^2} c_{p}) + t_{s}\sum_p  (c^\dagger_{p} s +s^\dagger c_{p})+  t_{1} \sum_{pk}(c^\dagger_{p} c_{k} + c^\dagger_{k} c_{p}),
H_s&=&\epsilon_s \sum_\sigma s^\dagger_\sigma s_\sigma + \sum_{\bk\sigma} V_{\bk A1s} (s^\dagger_\sigma c_{\bk A1\sigma}+ c_{\bk A1\sigma}^\dagger s_\sigma)+\nonumber\\
&&+t_{s}\sum_{\bp A1\sigma}  (c^\dagger_{\bp A1\sigma} s_\sigma +s_\sigma^\dagger c_{\bp A1\sigma})-\nonumber\\
&&-\boldsymbol{S}_s\cdot\sum_i J_{si}\boldsymbol{S}_i,
\eea
with on-site energy $\epsilon_s$, destruction operator $s_\sigma$ for spin $\sigma$, hybridization matrix elements $V_{\bk A1s}$ with $A1$ conduction states, coupling $t_s$ to $A1$ states of the tip, and exchange coupling $J_{si}$ with $d_{i}$ states. 
However, due do its large on ($\Gamma_s\sim 2$eV), the $4s$ orbital can be taken as part of the A1 conduction band, thus effectively enhancing $t_{1A1}$ in Eq.~\ref{h_tip_100} for the $d_{z^2}$ orbital.

In practice, since we are mainly interested in the Kondo temperature, instead of solving Eqs.~\ref{h_ad_sur100} and \ref{h_tip_100} together by NRG, we always solve Eq.~\ref{h_ad_sur100} alone but replace the hybridization $\Gamma_i^s$ due to the surface with the total hybridization $\Gamma_i=\Gamma_i^s+\Gamma_i^t$ due to the surface plus tip. 
%{\color{red} Where is it explained?: }
As explained in Sec. \ref{sec_ourapproach}, the lineshape is approximated % guessed
during the DFT step without resorting to the model Hamiltonian in Eqs.~\ref{h_ad_sur100} and \ref{h_tip_100}.

%, whose many parameters cannot all be determined unambiguously.

The parameters in Eq.~\ref{h_ad_sur100} are then fixed by trying to reproduce as closely as possible the GGA results within the HF approximation of the AIM in the wide-band limit \cite{lucignano},
which gives:
\bea
\epsilon_i^\uparrow=\epsilon_i+U_in_i^\downarrow+\sum_jU_{ij}n_j-\sum_jJ_{ij}m_j/4,\\
\epsilon_i^\downarrow=\epsilon_i+U_in_i^\uparrow+\sum_jU_{ij}n_j+\sum_jJ_{ij}m_j/4,
\eea
where we sum over all $j\ne i$ atomic orbitals ($n_i^\sigma$, with $\sigma=\uparrow,\downarrow$, is the occupation
of orbital $i$ in the spin channel $\sigma$; $n_i=n_i^\uparrow+n_i^\downarrow$; $m_i=n_i^\uparrow-n_i^\downarrow$).
The linewidths $\Gamma_{i}$ are taken from the down-spin density of states by fitting each impurity orbital with a Lorentzian; this actually gives $\Gamma_i(\epsilon_i^\downarrow)$, but we assume $\Gamma_i(\epsilon_i^\downarrow)\simeq \Gamma_i(0)$. $J$ is assumed to be constant in the $d$-shell. It is inferred from the energy splitting of the $d_{xy}$ orbital (which has a very low magnetization $m_{xy}\sim 0.04 \mu_B$), induced by the total magnetization $m$ via exchange interactions:
\be
J=\frac{2(\epsilon_{xy}^\downarrow-\epsilon_{xy}^\uparrow)}{m}.
\ee
Hubbard repulsions $U_i$ are taken from the splitting of magnetic orbital, once $J$ is known:
\be 
\epsilon_{i}^\downarrow-\epsilon_{i}^\uparrow=U_i m_i+\frac{J}{2}(m-m_i), \hspace{10pt}i=d_{z^2}, d_{x^2-y^2}.
\ee
The inter-orbital Hubbard repulsion $U_{12}$ is approximated by \cite{prb_nanotubi}
\be
U_{12} = U_{\rm ave}-\frac{5}{4}J {,}
\ee
where $U_{\rm ave}$ is the average of the Hubbard repulsion over the $d_{z^2}$ and $d_{x^2-y^2}$ orbitals. 
Finally, the on-site energies $\epsilon_i$ are fixed from the orbital energies $\epsilon_i^\sigma$ together with the knowledge of $U_i$ and $U_{12}$:
\be
\epsilon_i=\frac{\epsilon_i^\uparrow+\epsilon_i^\downarrow}{2}-\frac{U_i}{2}n_i-U_{12}n_j, \hspace{10pt}i,j=d_{z^2}, d_{x^2-y^2},j\neq i
\ee
Here, the numerical values of $\epsilon_i^\sigma$, $n_i^\sigma$, $n_i$ and $m_i$ are taken from DFT.  The results of this procedure is reported in Table~\ref{tab_cocu100}.

\subsection{Kondo temperature}\label{sec_tk_cocu100}
In order to estimate the Kondo temperature, Eq.~\ref{h_ad_sur100} for the $d_{z^2}$ and $d_{x^2-y^2}$ orbitals is solved by a two-channel NRG calculation. Each orbital, with on-site energy $\epsilon_i$ and Hubbard repulsion $U_i$, is coupled to its own Wilson chain through the full broadening $\Gamma_{i}=\Gamma_{i}^s + \Gamma_{i}^t$, which takes into account the interaction with both the surface and tip; the two channels are coupled by Hund exchange coupling $J$ and inter-orbital Hubbard repulsion $U_{12}$.
The Kondo temperature is obtained by computing the spectral function for both impurity levels,
and taking the halfwidth of the zero energy resonance.

GGA predicts the down-spin orbitals to lie just above the Fermi energy (see Fig. \ref{cocu100dos4}). 
When translated into an AIM, this means that $\epsilon_i^\downarrow \sim \epsilon_i + U_i \gtrsim 0$, which leads to high particle-hole asymmetry.
As a consequence, NRG predicts the magnetic orbitals to be almost in the mixed-valence region, which explains why the values of $T_K$ in Table~\ref{tab_cocu100} are higher than the experimental ones.  However, we believe this high particle-hole asymmetry is a spurious effect, which could be amended by resorting to some more sophisticated method, such as GGA+U \cite{lda+u}. For example, we found that a small value of $U\sim0.5$ eV in the GGA+U approach is enough to reproduce the experimental $T_K$ in the tunneling regime.

Our value of $\Gamma_{z^2}\simeq 0.18$ eV in the tunneling regime is comparable, but somewhat lower than other values found in the literature: 0.20 eV in Ref.~\onlinecite{ujsaghy} and 0.24 eV in Ref.~\onlinecite{Neel2007}.  
Even though our method of computing the hybridization from the broadening of scattering states is in principle more accurate than a simple estimation from the density of states after a self consistent calculation,
since it involves the interaction of impurity levels with a continuum of states, hence requiring no artificial broadening,
the approximation of using a single $\bk_{x,y}$ point can easily lead to a $\sim 10\%$ error,
as a consequence of the interaction among periodic replicas of the impurity in the $x$ - $y$ plane,
which  could be alternatively reduced by the use of larger supercells.
In addition, the way the nuclear relaxation is performed---either taking into account magnetic effects or ignoring them---is found to be another source of uncertainty.  
For example, performing a spin-unpolarized relaxation increases $\Gamma_{z^2}$ from 0.183 to 0.208 eV with respect to the standard spin-polarized calculations we use in this paper.  % This distinction was 
% claimed not to be a relevant issue
This differs with earlier work \cite{huang_cocu100}, where magnetic and nonmagnetic calculations were found to yield similar relaxed atomic coordinates due to a cancellation between adatom--substrate and adatom--tip interactions; see also Ref.~\onlinecite{Vitali2008}.  For comparison, we find the relaxed Co-surface distance without the tip to be $1.59\text{\AA}$ when magnetism is taken into account, and $1.49\text{\AA}$ otherwise; in Ref.~\onlinecite{huang_cocu100} this value was found to be $1.51\text{\AA}$, regardless of magnetism.

As such, a precise evaluation of the Kondo temperature remains a challenge. However, we stress that we obtain the correct growth of 
$T_{K,z^2}$ in passing from the tunneling regime to the contact regime, as reported in Table~\ref{tab_cocu100} and Fig.~\ref{tkplot}.  
In addition, in the intermediate regime, 
i.e.~for $g\sim 0.3$, $T_{K,z^2}$ decreases slightly, which is probably true experimentally (see Fig.~3a of Ref.~\onlinecite{choi_cocu100}). 
This is a consequence of the fact that the Co adatom is attracted by the tip, and therefore pulled away from the surface (see Table~\ref{tab_cocu100}, column $d_{Co-sur}$; this agrees well with Ref.~\onlinecite{huang_cocu100}).  
This causes a decrease in the hybridization $\Gamma_{z^2}$, which is compensated only in the contact regime, when the tip is close enough to the Co atom to be considered as an additional nearest-neighbor atom. 
At this point, $\Gamma_{z^2}$ and $T_{K,z^2}$ are greatly enhanced. However, this is only true for the $d_{z^2}$ orbital, which is probably the one producing the experimental ZBA.  
In the case of $d_{x^2-y^2}$, the tip greatly decreases $T_{K,x^2-y^2}$ when approaching the adatom, because the decreased hybridization of the orbital with the surface due to the increased surface--adatom distance is not compensated by an additional hybridization with the tip, due to symmetry mismatch; should symmetry be broken, things would not be different, since the $d_{x^2-y^2}$ orbital lies flat on the surface, thus hardly coupling with the STM tip no matter how this approaches the adatom.  
This is true until the tip really ``pushes'' the atom closer to the surface, eventually enhancing $T_{K,x^2-y^2}$ too. 
We stress that, in contrast to Refs. \onlinecite{choi_cocu100, Neel2007} we attribute the change of the Kondo temperature mainly to changes in the hybridization, rather than in the on-site energy and Hubbard repulsion.

\subsection{Lineshape}
As remarked, in our symmetry-preserving geometry, the total lineshape is the sum of the $A1$ and $B1$ conductances in Eq.~\ref{gtot100}. 
However, after solving Eq.~\ref{h_ad_sur100} with the parameters shown in Table~\ref{tab_cocu100}, we find that $T_{K,z^2}\gg T_{K,x^2-y^2}$; moreover, $g_{A1}\gg g_{B1}$ because the $d_{x^2-y^2}$ orbital is flat and only couples to the second layer of the tip for symmetry reasons. 
This suggests that most of the experimental Kondo signal is due to the $d_{z^2}$ orbital and, in what follows, we shall assume $g_{tot}=g_{A1}$ and ignore $g_{B1}$, together with all other symmetry channels, which do not carry ZBA's and contribute very little to the conductance. In any case, we find that the Fano parameter associated with the orbital $d_{x^2-y^2}$ should always be much larger than 1, %, $q_{x^2-y^2}\gg 1$, 
and this would show up as an additional weak anomaly superimposed to the standard one.

Applying the phase-shift analysis described in Ref.~\onlinecite{lucignano} turned out to be too cumbersome for this system,
due to its intrinsic 3D character, so instead we estimated $q$ 
% we made the approximation of looking at 
from the shape of the energy-dependent DFT transmission coefficient, which shows interference at energies around $\epsilon_d^\downarrow \sim 0.5$ eV. 
If we assume that the hopping parameters in Eq.~\ref{h_tip_100} are weakly energy dependent, the DFT lineshape can be a good approximation to the ZBA, which involves interference at the Fermi energy. 
Unfortunately, when we do so, the agreement with experiment is not always good. 
In the tunneling regime, we obtain $q\sim 1$, which nicely matches experiments \cite{knorr2002}. 
However, when going into the contact regime, we find a decrease of $q$. 
In this regime the conductance $G$ is close to the unitary value $G_0$, thus the interference between the $s$ and $d_{z^2}$ orbitals can only be destructive, leading to a dip in the conductance, while in experiments the opposite is observed ($q$ increases slightly \cite{Neel2007} or strongly \cite{choi_cocu100}). 

A possible reason for this disagreement in the contact regime is 
that the junction is formed in a different way than we have modeled it.  
%non-equilibrium effects start to play a decisive role, the lack of symmetry play a much larger role than we expect, or 
Other sources of error in our calculations are the inclusion of non-equilibrium effects and our assumption of energy-independent parameters; see also the conclusions in Sec. \ref{sec_conclusions}.

\section{Co/Cu(111)}\label{sec_cocu111}
In this section we report our results for Co on the Cu(111) surface, which has $T_K=54$ K and $q=0.18$ in the tunneling regime \cite{knorr2002}. These values remain almost unchanged in passing to the contact regime \cite{Vitali2008}.

\subsection{DFT results}
Once again, we find that Co adsorbs in the hollow position, this time with 3 nearest-neighbor Cu atoms in a configuration with $C_{3v}$ symmetry.  The $d$ orbitals split into a singlet with symmetry $A_1$ ($d_{z^2}$) and two doublets with symmetry $E$ ($d_{\alpha 1,2}$, $d_{\beta 1,2}$); the $s$ orbital has A1 symmetry. The electronic configuration is $3d^84s^1$, and the total magnetic moment is close to 2$\mu_B$, just like on the (100) surface.
We again model the tip to preserve the symmetry ($C_{3v}$).

\begin{figure}[htb]
\centering
 \includegraphics[scale=0.28]{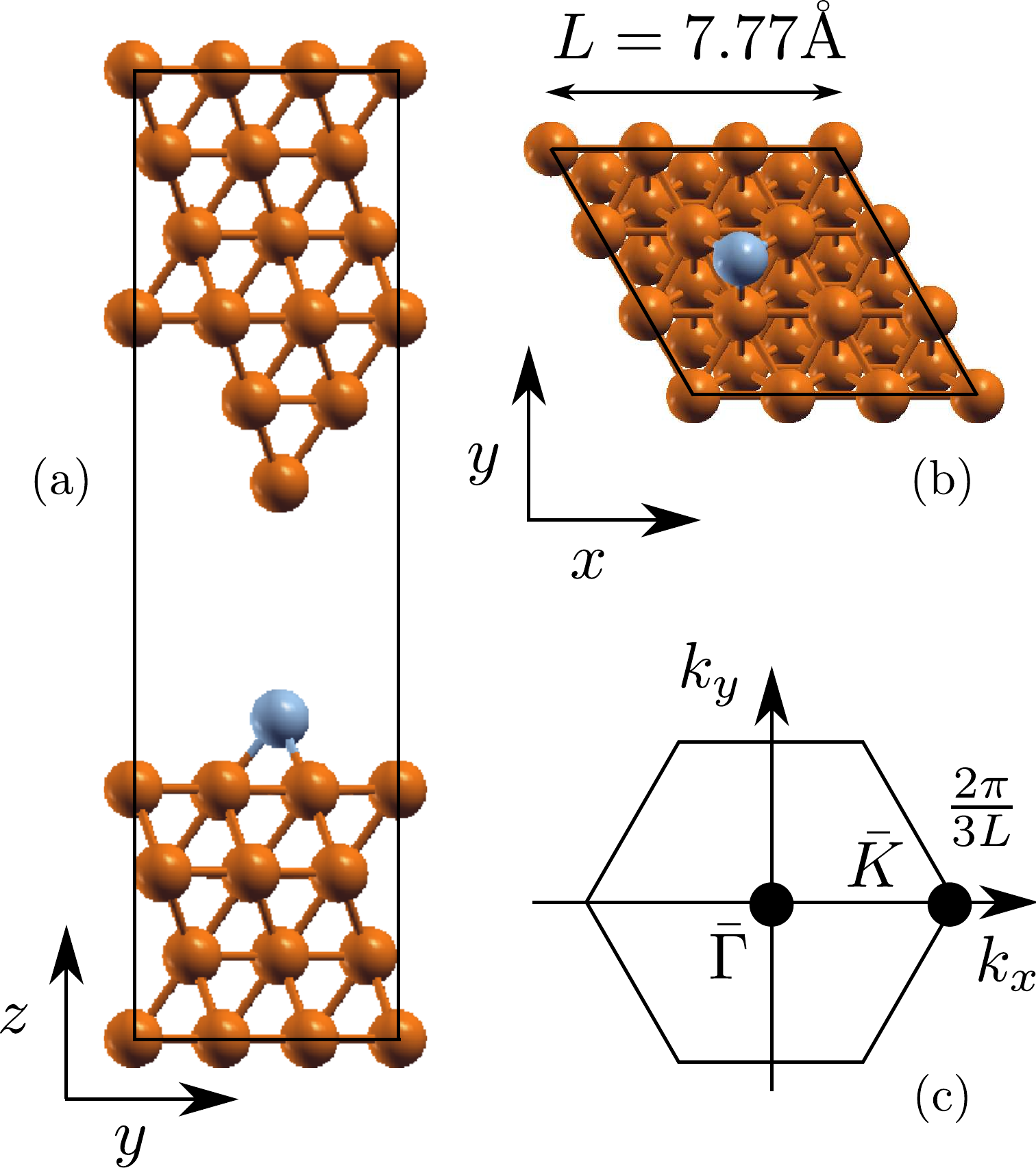}
\caption{Unit cell of the scattering region for Co/Cu(111) for $d_{tip-sur}=7.81\text{\AA}$ (a) in the $yz$ plane and (b) in the $xy$ plane. (c) 
Brillouin zone for the supercell in the $xy$ plane.
%Brillouin zone in the $\bk_{x,y}$ plane for the supercell.
}\label{cocu111cell}
\end{figure}

%Due to symmetry reasons, 
If we take Cartesian axes as in Fig. \ref{cocu111cell}, the doublets can be written in the following way:
\bea
d_{\alpha 1}&=&\cos\theta d_{xz} + \sin\theta d_{xy},\\
d_{\alpha 2}&=&\cos\theta d_{yz} + \sin\theta d_{x^2-y^2},\\
d_{\beta 1}&=&-\sin\theta d_{xz} + \cos\theta d_{xy},\\
d_{\beta 2}&=&-\sin\theta d_{yz} + \cos\theta d_{x^2-y^2},
\eea
where the $d_{\alpha 1,\beta 1}$ orbitals are odd with respect to the symmetry operator $P_x: x\rightarrow -x$, while the $d_{\alpha 2,\beta 2}$ orbitals are even. From DFT calculations, it turns out that the $d_{\alpha 1,2}$ doublet is magnetic, while $d_{\beta 1,2}$ is fully occupied; moreover we find $\theta=0.70$ rad.

\begin{figure}[ptb]
 \includegraphics[width=0.5\textwidth]{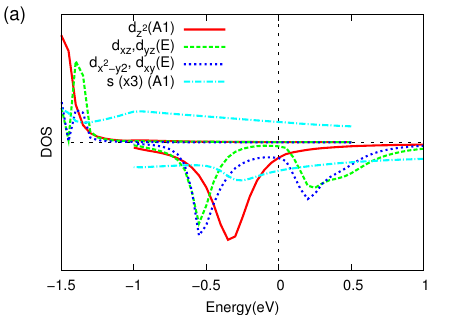}
 \includegraphics[width=0.5\textwidth]{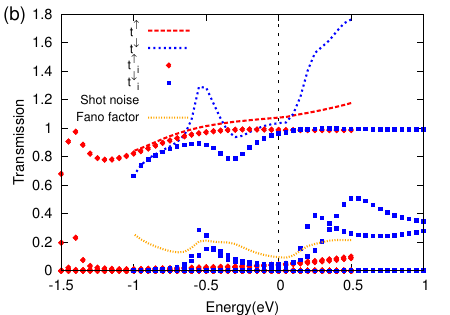}
\caption{Co/Cu(111): (a) spin-polarized density of states { at Co $3d$ and $4s$ atomic orbitals constructed from scattering states} at $\bar{K}=\frac{\pi}{L}(\frac{2}{3},0)$ (positive DOS means spin up, negative DOS spin down); 
(b) { $t^{\uparrow}$ and $t^{\downarrow}$ components of the DFT transmission, 
transmission eigenvalues $t^{\uparrow}_i$ and $t^{\downarrow}_i$ 
(eigenvalues close to 1 correspond to the A1 channel while two lower eigenvalues are from the E channels),} 
and the shot noise Fano factor.  
Energies are with respect to the Fermi energy. 
}\label{cocu111dos252} 

\end{figure}

In Table~\ref{tab_cocu111} we show some structural and electronic data for different tip--surface separations.  In Fig.~\ref{cocu111dos252} we show density of states and transmission eigenvalues for the shortest distance, $d_{tip-sur}=4.33\text{\AA}$.  At $\bar{K}$, the degeneracy between $d_{\alpha 1}$ and $d_{\alpha 2}$ is weakly broken, so two different peaks appear in the density of states and in transmission eigenvalues at $\sim -0.5$eV ($d_{\beta}$ orbitals) and $\sim 0.5$eV ($d_{\alpha}$ orbitals) for down-spin electrons in the E channels.

\begin{table*}
\centering
\begin{ruledtabular}
\begin{tabular}{cccccccccccccc}
$d_{tip-sur}$ & $d_{tip-Co}$ & $d_{Co-sur}$ & $d_{Co-nn}$ & $m$ & $\epsilon_{\alpha}$ & $U_{\alpha}$ & $\Gamma_{\alpha}$ & $U_{12}$ & $J$ & $T_K$ & $q$ & $g$ & $F$ \\\colrule
%6.00	
7.81	&6.01	&1.80	&2.40&2.20
&-5.20&3.26&0.164&1.90&1.09&70&7.1&0.01&0.99\\
%4.00	
5.81&	3.92	&1.89&2.44&2.20
&-5.33&3.34&0.168&1.98&1.09&50&6.4&0.47&0.53\\	
%3.60	
4.33&	2.44&	1.89&	2.44&	2.19
&-5.39&3.28&0.157&1.92&1.09&100&4.5&1.06&0.10
%3.15	
\end{tabular}
\end{ruledtabular}
\caption{Parameters for Co/Cu(111) at different tip--surface separations $d_{tip-sur}$: 
the distance $d_{tip-Co}$ between the tip and the Co adatom, 
$d_{Co-sur}$ between the Co adatom and the surface, 
$d_{Co-nn}$ between the Co adatom and its 3 nearest neighbors, 
the total magnetization $m$ of the unit cell in units of $\mu_B$, 
the on-site energy $\epsilon_\alpha$, 
the Hubbard repulsion $U_\alpha$, 
the hybridization $\Gamma_\alpha$, 
the inter-orbital Hubbard repulsion $U_{12}$, 
the Hund exchange $J$, 
the calculated Kondo temperature $T_K$ in $K$,
the Fano parameter $q$,
the DFT { conductance} $g=g^\uparrow+g^\downarrow$ and the Fano factor F at the Fermi energy; distances are in \mbox{\AA} and
energies in eV.}\label{tab_cocu111}
\end{table*}

% from the NRG solution of Eq.~\ref{h_ad_sur111}


\subsection{Anderson model}
The full atomic Hamiltonian is
\bea
H&=&\sum_{i=z^2, \alpha 1, \alpha 2,\beta 1, \beta 2 } (\epsilon_i n_i +U_i\,n_{i\up}\,n_{i\down})-\nonumber\\
&&-\sum_{j< i}\,J_{ij} \,\boldsymbol{S}_i\cdot\boldsymbol{S}_j+\sum_{j<i}\, U_{ij} n_i n_j,
\eea
with on-site energies $\epsilon_{\alpha 1}=\epsilon_{\alpha 2}\equiv \epsilon_{\alpha}$, $\epsilon_{\beta 1}=\epsilon_{\beta 2}\equiv \epsilon_{\beta}$, Hubbard repulsion $U_{\alpha 1}=U_{\alpha 2}\equiv U_{\alpha}$, and $U_{\beta 1}=U_{\beta 2}\equiv U_{\beta}$, inter-orbital Hubbard repulsion $U_{ij}$  and Hund exchange coupling $J_{ij}$. After dropping fully-occupied orbitals, which means keeping only $d_{\alpha 1}$ and $d_{\alpha 2}$, we introduce metallic states, to get 
\bea\label{h_ad_sur111}
H_{ad-sur}&=&\sum_{\substack{\bk\sigma\\i=\alpha 1,\alpha2}} 
\epsilon_{\bk}c^\dagger_{\bk i\sigma}c_{\bk i\sigma}+\nonumber\\
&&+\sum_{\substack{\bk\sigma\\i=\alpha1,\alpha2}} V_{k\alpha} (c^\dagger_{\bk i\sigma} d_{i\sigma}+d_{i\sigma}^\dagger c_{\bk i\sigma} )+\nonumber\\
&&+\sum_{i=\alpha1,\alpha2}(\epsilon_\alpha n_i+ U_\alpha n_i^\uparrow n_i^\downarrow)+ \nonumber\\
&&+U_{12} n_{\alpha1}n_{\alpha2}-J\boldsymbol{S}_{\alpha1}\cdot \boldsymbol{S}_{\alpha2}.
\eea
%where 1 stands for $\alpha 1$ and the associated conduction band, and 2 for $\alpha 2$ and its conduction band.

In addition to the $d$ orbitals, the $s$ orbital is half-filled and highly hybridized, exactly as on the (100) surface, with the same Hamiltonian (Eq.~\ref{h_s}),
%\be
%H_s=\epsilon_s n_s + V_s \sum_k (c_k^\dagger s + s^\dagger c_k)-  \boldsymbol{\sigma}_s\cdot\sum_i J_{si}\boldsymbol{\sigma}_i,
%\ee
but it is irrelevant when dealing with Kondo physics: it only contributes to the conductance in the $A1$ channel which shows no ZBA, and does not interfere with the $d_\alpha$ orbitals, which have different symmetry $E$.

Once again, a generalized Tersoff-Hamann model 
\bea
H_{tip}&=&\sum_{\substack{\bp\sigma\\i=\alpha1,\alpha2}} \epsilon_{\bp i} c^\dagger_{\bp i\sigma} c_{\bp i\sigma} + \\
&&\sum_{\substack{\bp\bk\sigma\\i=\alpha1,\alpha2}} t_{1i} (c^\dagger_{\bp i\sigma} c_{\bk i} + c^\dagger_{\bk i\sigma} c_{\bp i\sigma}) +\\
&&\sum_{\substack{\bp\sigma\\i=\alpha1,\alpha2}} t_{2i} (c^\dagger_{\bp i\sigma} d_{i\sigma}+d^\dagger_{i\sigma} c_{pi\sigma}),
\eea
is used and the hybridization functions
\bea
\Gamma_{\alpha}^s(\epsilon)&=&
\pi\sum_\bk \delta (\epsilon-\epsilon_\bk) V_{\bk\alpha}^2\rightarrow\pi\rho_s V_{\alpha}^2,\\
\Gamma_{\alpha}^t(\epsilon)&=&
\pi\sum_\bk \delta (\epsilon-\epsilon_\bk) t_{\bk\alpha}^2\rightarrow\pi\rho_t t_{2\alpha}^2,\\
\Gamma_{\alpha}(\epsilon)&=&\Gamma_{\alpha}^s(\epsilon) + \Gamma_{\alpha}^t (\epsilon) 
\eea
are introduced and approximated as energy-independent quantities $\Gamma^s_{\alpha}$, $\Gamma^t_{\alpha}$, $\Gamma_{\alpha}$.
%where 1 stands for orbital $\alpha_1$ and the associated band, 2 for $\alpha_2$ and the associated band, both of symmetry E, and 3 stands for orbital $s$ and symmetry $A1$; $t_{11}=t_{12}$, $t_{21}=t_{22}$.

The parameters are fixed in the same way as for the Co/Cu(100) case; we get $D_{\alpha}=5$ eV from the density of states of surface $s$ orbitals; $J$ is fixed from the splitting of $\beta$ orbitals, which have $m_\beta=0.06$. Numerical values are reported in Table~\ref{tab_cocu111}.  The value $\Gamma_{\alpha}\sim 0.16$ eV is slightly below the value $0.18$ eV of Ref. \onlinecite{castroneto_cocu111}.

\subsection{Kondo temperature}
The Kondo temperature is obtained by solving Eq.~\ref{h_ad_sur111} with NRG for the model parameters in Table~\ref{tab_cocu111}.  We used the total hybridization $\Gamma_{\alpha}$, including contributions from both the surface and tip.

Since the $d_{\alpha1}$ and $d_{\alpha2}$ orbitals are degenerate, there is a single Kondo temperature as reported in Table~\ref{tab_cocu111}. Spin-orbit effects will lift this degeneracy, leading to two different Kondo temperatures, but the effect should be small. As for the previous case, we overestimate the Kondo temperature due to the excessive particle-hole asymmetry which comes from GGA, but in this case the disagreement is not too bad.

When the tip is brought down to the surface, we find that the Kondo temperature first decreases slightly (even though the difference is below the accuracy of our method), because the adatom is pulled farther away from the surface, then weakly increases.  The $s$ orbital of the tip is of different symmetry than the $\alpha$ orbitals, so $\Gamma_\alpha^t=0$ to a first approximation.  The effect of the tip-induced relaxation 
of the Co adatom and its neighbors is found to be negligible, in contrast to the (100) case, so $\Gamma^s_\alpha$ is basically unaffected by the tip. This is in good agreement with a similar analysis in Ref.~\onlinecite{Vitali2008} and with experimental results that show a constant $T_K$ as a function of the tip position \cite{Vitali2008}. At a distance $d_{tip-sur}=4.33\mbox{\AA}$, the Co atom is pushed towards the surface, and at this point the Kondo temperature starts to increase, but this regime is probably not reached in experiments.

\begin{figure*}[ptb]
\includegraphics[width=0.45\textwidth]{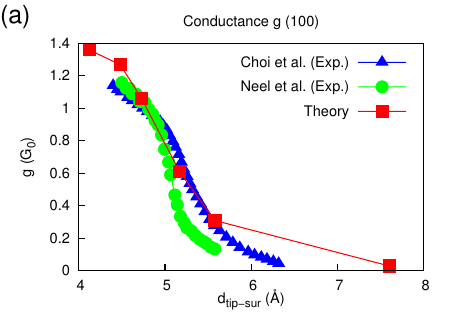}
\includegraphics[width=0.45\textwidth]{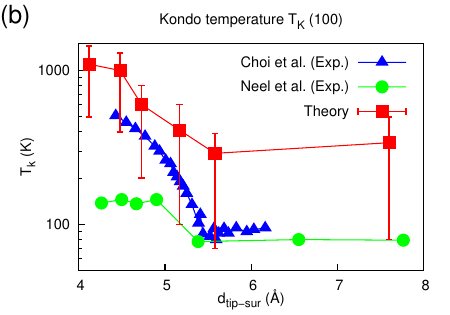}
\includegraphics[width=0.45\textwidth]{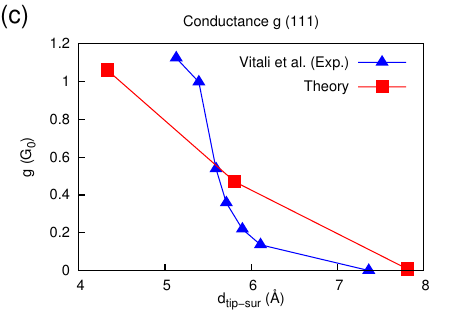}
\includegraphics[width=0.45\textwidth]{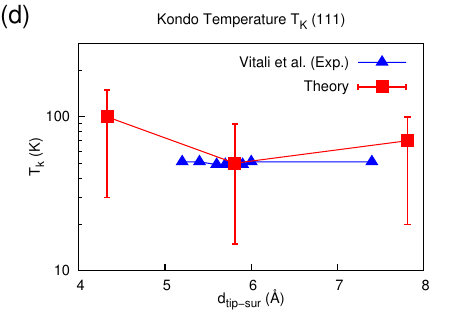}
\caption{
%(a): hybridizations $\Gamma_i$ in eV, (b): Kondo temperatures $T_{K,i}$ in K for Co orbitals $d_{z^2}$ and $d_{x^2-y^2}$ on Cu(100) and for Co orbitals $d_{\alpha}$ on Cu(111), 
(a): 
{ DFT conductance $g$} and (b): Kondo temperature $T_K,z^2$ on the Cu(100) surface compared with experimental data from Refs. \onlinecite{choi_cocu100} (Choi et al.) and \onlinecite{Neel2007} (Neel et al.), 
(c): {DFT conductance $g$} and (d): Kondo temperature on the Cu(111) surface compared with experimental data from Ref. \onlinecite{Vitali2008} (Vitali et al.).
All data are shown as a function of the distance between the tip and the surface $d_{tip-sur}$: since only relative distances are known experimentally, we have rigidly shifted experimental data by 5~\AA~for Ref.\onlinecite{choi_cocu100}, 9~\AA~for Ref. \onlinecite{Neel2007} and 5.5~\AA~for Ref. \onlinecite{Vitali2008} for best fit with theory. Approximate error bars are shown for the theoretical Kondo temperatures: they take into account all the incertitudes of our method when fixing the parameters of the Anderson Hamiltonian Eqs. \ref{h_ad_sur100}, \ref{h_ad_sur111} from the DFT calculations; once the parameters are known, the Kondo temperature
can be extracted with negligible uncertainty from the NRG run. It must be stressed that most of the incertitudes are systematic, so they affect all the data in the same direction, which turns out, when compared to experiments, to be an overestimation of the Kondo temperatures.
%The reported values of $T_{K,i}$ are to be seen mostly as upper estimates, since the use of GGA+U would in any case lead to a decrease of all Kondo temperatures; a small increase of $T_{K,i}$ can come from the incertitude in estimating the hybridizations $\Gamma_i$. 
 }\label{tkplot}
\end{figure*}

In Fig.~\ref{tkplot} we show the evolution of the conductance g and Kondo temperature $T_{K,i}$ as a function of the tip--surface distance for both Co/Cu(100) and Co/Cu(111).  The reported values of $T_{K,i}$ are to be seen mostly as upper estimates, since the use of GGA+U, as mentioned, would decrease the Kondo temperature.  In fact, our values overestimate the experimental Kondo temperature.
%!
% Uncertainties in $\Gamma_i$ might cause a small increase of $T_{K,i}$ can come from . 
It must be stressed that most of the uncertainties are systematic, so they affect all the data in the same direction.
%, which turns out, when compared to experiments, to be an overestimation of the Kondo temperatures. 

\subsection{Lineshape}
When the tip is above the adatom, symmetry is preserved, and the total conductance is 
\be
{ g_{tot}= \sum_{i=A1,\alpha,\beta} g_i \simeq 2g_\alpha + G_{A1}.}
\ee
In the A1 channel, there is no ZBA, because only the $s$ orbital is involved, the  $d_{z^2}$ orbital being completely filled.  
In the $E$ channel, there is a ZBA due to $d_{\alpha}$ orbitals, but the signal should be small because they do not couple to the $s$ orbital of the tip.  
DFT predicts that these channels give a peak in the conductance ($q \gg 1$, see Fig. \ref{cocu111dos252}), the coupling of the tip to the orbitals being much higher than to Cu states, according to the two-path model, but still much lower than the coupling to the Co $s$ orbital. 
However, when compared to experiment, where a dip ($q\sim 0$) in conductance is observed, with a strong signal, this is wrong. 
Moreover,  
on the basis of the DFT results, one would expect to see considerable changes in the lineshape when moving the tip laterally in the $xy$ plane, due to symmetry breaking, but this is not seen either.  
Finally, when the tip approaches the surface, the lineshape is unaffected, remaining a minimum, 
again in contradiction with what one would expect from the DFT results.

This probably means that symmetry is unimportant, most likely because the tip breaks it. This implies that one can observe $E$ orbitals even when the tip is above the adatom, and when moving the tip laterally there is no symmetry breaking and no significant change in the signal. However, this is not sufficient to explain the experimental results because one would still expect the ZBA to be a peak. %maximum
Another interpretation might be that since the magnetic orbitals have $E$ symmetry, while tunneling only happens through metal states of $A1$ symmetry, these acquire a dip in their density of states, leading to $q\sim0$, as suggested in Ref.~\onlinecite{huang_carter_2008}; however, metal states of $A1$ symmetry should carry no, or at most very weak, Kondo signal. 

Instead, it is likely that surface states are responsible for the dip. 
%playing a major role.  
According to DFT, the tip ``sees'' the magnetic orbital and not the metallic states, while it should be the other way around. 
This is simply because the Co adatom is closer to the tip than to the Cu surface atoms (on the (100) surface things are different, because the Co $s$ orbital, which is effectively part of the conduction band, can interfere with the magnetic orbital.) However, surface states, if taken into account, might prove to be more prominent than the $d$ orbitals, leading to a dip in conductance.  
This interpretation is also suggested by the ``quantum mirage'' \cite{Manoharan_mirage} experiment, which shows how surface states can carry Kondo information even far from the impurity, with the same lineshape as when the tip is above the adatom. 
Our DFT slab calculation can in principle include surface states, but to describe them correctly we would need a much larger supercell.

The role of surface states has been discussed in the literature.  
The total hybridization from the surface $\Gamma^s=\Gamma_{surf}+\Gamma_{bulk}$ is the sum of the hybridization from surface states and bulk states.  
While it is generally agreed \cite{castroneto_cocu111, barral_surfstates, knorr2002, wahl_coag, merino_gunnarsson, cornaglia_balseiro}, the only exception being Ref.~\onlinecite{henzl_surfacestates}, based on calculations and experimental hints (for example the lack of appreciable changes in the Kondo temperature at step edges and defects \cite{limot_surfacestates}, where surface states are deeply affected),
% deeply affect surface states, do not change the Kondo temperature appreciably  
that $\Gamma_{bulk}$ is much larger than $\Gamma_{surf}$, by up to a factor of one hundred \cite{castroneto_cocu111}, 
not much is known about the relative magnitude of tip--surface and tip--$d$ orbitals coupling, which is what controls the lineshape. 
It is only known that for a clean surface about two thirds of the current flows into surface states \cite{jeandupeux_surfacestates}.  Ref.~\onlinecite{merino_gunnarsson_prl} argues that surface states can give an important contribution to the conductance even in the presence of adsorbates.

In any case, it must be stressed that the usual assumption that the magnetic orbital is of $d_{z^2}$ character is found to be wrong in this case.
This makes the usual Tersoff-Hamann approach fail, because due to symmetry mismatch there can be no conductance from the $s$-state of the tip into the magnetic orbital, or to the linear combination of conduction states to which the magnetic orbital is coupled, which leads to the paradox that when the tip is directly above the adatom, it should give almost no signal.

\section{Conclusions}\label{sec_conclusions}
We have demonstrated the application of our DFT+NRG method \cite{lucignano} to Co adatoms on Cu surfaces probed by an STM tip, trying to describe the lineshape and the Kondo temperature both in the tunneling and contact regimes.

The calculations show the severe difficulty in predicting the details of experimental Kondo anomalies, especially as far as the lineshape is concerned, while the trend in the Kondo temperature, if not its absolute value, is reliable.  %rationalized. 
Several different issues might cause these discrepancies.

First of all, our description of the tip is surely oversimplified.  Yet, experimental measurements are only weakly dependent on the choice of the tip, at least in the tunneling regime, so in principle this should not be a large source of error.

We believe that on the Cu(100) surface the essential remaining point is to correct the estimation of the parameters which control the lineshape: the asymmetric ZBA in the tunneling regime can be understood in terms of the interference between the $s$ and $d_{z^2}$ orbitals of the Co adatom. Issues may arise in the contact regime due to the geometrical details of the contact, which may differ from our model. The momentum dependence of the hopping parameters such as $V_\bk$, $t_{1i\bp\bk}$, $t_{2i\bp}$, which we have ignored, might also play a role, as well as non-equilibrium effects. Also, in the contact regime, which is not far from a bulk impurity situation, all $d$ orbitals start to become magnetized, thus actively entering conduction processes, and making our two-orbital model insufficient.

On the Cu(111) surface, in contrast, more work should be done to include surface states.  This might reverse the sign of the ZBA (giving a dip instead of a peak). Also, one should seek to understand if the $E$ symmetry of the magnetic orbital, instead of $A1$ as usually assumed, can affect the lineshape, as suggested e.g.~in Ref.~\onlinecite{huang_carter_2008}.

On both surfaces our estimate of the Kondo temperature would be improved by correcting the excessive particle-hole asymmetry brought about by plain GGA, for example by using GGA+U \cite{lda+u}.  
% by reaching convergence for the broadening $\Gamma_i$, which appears to be somewhat lower here than in other approaches. In our method, this would require either a larger supercell, or a greater number of $\bk_{x,y}$ points for which the conductance is calculated. 
Moreover, our many-body model may be improved.  For example, one could take into account spin-orbit effects, correlated hopping, double hopping, and other two-body interactions, or keep all $d$ orbitals, considering that some of them are almost but not completely filled, especially in the contact regime.  
In addition, the energy dependence of the hybridization functions $\Gamma_i(\epsilon)$ is likely to have some impact.  % play a role
{Finally, we note that the co-existence of many magnetic solutions (not only of lowest energy presented in this paper) 
realized at different Co adatoms (on both Cu surfaces ) could be responsible for statistical spread in Kondo temperatures
and lineshapes.}

We emphasize that, according to our GGA results, Co has spin $S=1$ on both Cu surfaces, showing that the usual assumption of $S=1/2$ is probably wrong. 
However, this mistake might not have a big impact on the final result if one of the two magnetic orbital has a much lower Kondo temperature than the other one, as on the (100) surface, or if the two magnetic orbitals are degenerate, as on the (111) surface.  Of course, a detailed quantitative approach cannot overlook this fact.

In any case, despite several works claiming that the Kondo physics of adatoms is fully understood, we believe that further effort is needed to completely understand, or at least describe satisfactorily from first principles, the Kondo behavior of Co adatoms on Cu surfaces, and, more generally, of magnetic adatoms on metallic surfaces.

\begin{acknowledgments}
This work was supported by PRIN/COFIN  20087NX9Y7 and under the ERC Advanced Grant No. 320796-MODPHYSFRICT. We acknowledge useful discussions with R. Zitko.
\end{acknowledgments}

\bibliographystyle{apsrev4-1}
\bibliography{biblio.bib}

\end{document}